\documentclass[aps,prl,twocolumn,showpacs,floatfix,superscriptaddress,letterpaper,longbibliography]{revtex4-1}
\usepackage{graphicx}
\usepackage{bm}
\usepackage{amsmath,amsfonts}
\usepackage{amsbsy}
\usepackage[utf8]{inputenc}
\DeclareUnicodeCharacter{03BC}{\mbox{$\mu$}}
\DeclareUnicodeCharacter{2009}{ }

\begin{document}

\title{Second sound in systems of one-dimensional fermions}

\author{K. A. Matveev}

\affiliation{Materials Science Division, Argonne National Laboratory,
  Argonne, Illinois 60439, USA}

\author{A. V. Andreev}

\affiliation{Department of Physics, University of Washington, Seattle,
  Washington 98195, USA}

\date{December 18, 2017}

\begin{abstract}

  We study sound in Galilean invariant systems of one-dimensional
  fermions.  At low temperatures, we find a broad range of frequencies
  in which in addition to the waves of density there is a second sound
  corresponding to ballistic propagation of heat in the system.  The
  damping of the second sound mode is weak, provided the frequency is
  large compared to a relaxation rate that is exponentially small at
  low temperatures.  At lower frequencies the second sound mode is
  damped, and the propagation of heat is diffusive.

\end{abstract}

\maketitle

The low-energy properties of systems of one-dimensional interacting
fermions are usually described in the framework of the
Tomonaga-Luttinger liquid theory \cite{tomonaga_remarks_1950,
  luttinger_exactly_1963, haldane_luttinger_1981,
  giamarchi_quantum_2004}.  Its main feature is that the elementary
excitations of the system are treated as noninteracting bosons with
linear dispersion.  The advantage of this approach is that it
adequately describes the low-energy properties of the system at any
strength of interaction between the fermions.  This theory provided
the foundation for understanding the basic properties of
one-dimensional electron systems, such as the power law
renormalizations of the impurity scattering and tunneling density of
states \cite{kane_transmission_1992, furusaki_single-barrier_1993},
observed in subsequent experiments \cite{tarucha_reduction_1995,
  auslaender_experimental_2000, bockrath_luttinger-liquid_1999,
  yao_carbon_1999}.

Much of the recent work on the theory of one-dimensional systems
focused on the properties not captured by the Luttinger liquid
picture, such as the nature and lifetimes of elementary excitations in
these systems.  When the interactions between the bosonic excitations
are taken into account, the excitations in spinless Luttinger liquids
become fermions \cite{rozhkov_fermionic_2005} with finite decay rate
$\tau_{\rm ex}^{-1}\propto T^\gamma$, with the exponent $\gamma=7$
\cite{imambekov_one-dimensional_2012, arzamasovs_kinetics_2014,
  protopopov_relaxation_2014} or 6
\cite{ristivojevic_relaxation_2013}, depending on the details of the
interaction between the physical particles forming the Luttinger
liquid.  For weakly interacting spin-$\frac12$ fermions $\tau_{\rm
  ex}^{-1}\propto T$ \cite{karzig_energy_2010}.  Importantly, the
scattering processes giving rise to the decay of elementary
excitations do not involve backscattering of fermions, i.e., each
quasiparticle remains in the vicinity of the nearest Fermi point.  The
backscattering processes involve hole states near the bottom of the
band, and their rate is exponentially small, $\tau^{-1}\propto
e^{-D/T}$ \cite{lunde_three-particle_2007, micklitz_transport_2010,
  matveev_equilibration_2010, matveev_equilibration_2012,
  matveev_scattering_2014}, where $D$ is the energy scale of the order
of Fermi energy.

In this paper we consider the dynamics of a system of one-dimensional
fermions in the absence of disorder at low temperatures $T\ll D$.
Such a system possesses three conserved quantities: the total number
of particles $N$, energy $E$, and momentum $P$.  At very low
frequencies $\omega\ll\tau^{-1}$ the system is close to equilibrium and
can be described by classical hydrodynamics.  We will be primarily
interested in the regime
\begin{equation}
  \label{eq:frequency_range}
  \tau^{-1} \ll \omega \ll \tau_{\rm ex}^{-1}.
\end{equation}
In this case the gas of elementary excitations is in thermal
equilibrium, but can move with velocity $u_{\rm ex}$ not equal to the
velocity $u$ of the center of mass of the fluid
\cite{matveev_equilibration_2012}.  At such frequencies the system
possesses a fourth conserved quantity: the difference between the
numbers of the right- and left-moving fermions $J=N^R-N^L$.  Because
relaxation of $J$ involves backscattering of fermions, it is
negligible at $\omega\gg\tau^{-1}$.

The detachment of the gas of elementary excitations from the rest of
the fluid is a well-known feature of superfluid $^4$He
\cite{landau_theory_1941, khalatnikov_introduction_2000}.  The
appropriate theoretical description of the motion of this system is in
terms of two-fluid hydrodynamics that predicts the existence of two
sound modes.  The first sound is the usual wave of
particle density, whereas the second sound is a wave of entropy that
propagates at a different velocity.  Our goal is to develop a similar
two-fluid hydrodynamics of the system of one-dimensional fermions in
the frequency range (\ref{eq:frequency_range}) and to demonstrate the
existence of the second sound in this system.

We will focus on the system of one-dimensional spin-$\frac12$ fermions
of mass $m$ with repulsive interactions and assume spin rotation
symmetry and Galilean invariance.  To leading order in $T/D\ll 1$, the
dynamics of the system is described by the conventional Luttinger
liquid theory with linear excitation spectrum \cite{footnote}.  Our
system supports two branches of bosonic excitations, corresponding to
the charge and spin sectors of the Hamiltonian, and propagating at
different velocities, $v_\rho$ and $v_\sigma$.  The momentum of the
system is \cite{haldane_luttinger_1981}
\begin{equation}
  \label{eq:P}
  P=\frac{h}{4L}NJ+\sum_kk(N^\rho_{k}+N^\sigma_{k}),
\end{equation}
where $N$ is the total number of fermions in a system of size $L$ with
periodic boundary conditions, and $h$ is the Planck constant, while
$N^\rho_{k}$ and $N^\sigma_{k}$ are the occupation numbers of the
bosonic excitations with momentum $k$ in the charge and spin channels,
respectively.  The first term in Eq.~(\ref{eq:P}) accounts for the
fact that at $N^R\neq N^L$ the ground state of the system has a
nonvanishing momentum $p_FJ$, where the Fermi momentum $p_F=hN/4L$.
Similarly, the energy of the system is given by
\begin{equation}
  \label{eq:E}
  E=\frac{mv_\rho^2}{2N_0}(N-N_0)^2+\frac{h^2NJ^2}{32mL^2}
    +\sum_k[\epsilon_\rho(k)N^\rho_{k}+\epsilon_\sigma(k)N^\sigma_{k}],
\end{equation}
c.f.~\cite{haldane_luttinger_1981}.  In the first term $N_0$ is some
reference value of the particle number and we have used the usual
relation between the ground state compressibility and $v_\rho$.
Bosonic excitations in the Luttinger liquid are superpositions of
small momentum particle-hole pairs near each Fermi point.  At
$N^R=N^L$ the energies are
$\epsilon_{\rho,\sigma}(k)=v_{\rho,\sigma}|k|$.  At $N^R\neq N^L$ the
quasiparticle ground state is moving with velocity
\begin{equation}
  \label{eq:u_0}
  u_0=\frac{hJ}{4mL}.
\end{equation}
The dependence of the  quasiparticle energies on $u_0$,
\begin{equation}
  \label{eq:epsilon}
  \epsilon_{\rho,\sigma}(k)=v_{\rho,\sigma}|k|+u_0k,
\end{equation}
is obtained by performing Galilean transformation to the stationary
frame. 

At frequencies below $\tau_{\rm ex}^{-1}$, collisions between the
bosonic excitations occur very quickly compared with the typical time
scale $\omega^{-1}$, and to first approximation one can assume that
the gas of excitations is in an equilibrium state described by the
Bose distribution
\begin{equation}
  \label{eq:Bose}
  N^{\rho,\sigma}_{k}=\left[\exp\left(\frac{
        \epsilon_{\rho,\sigma}(k)-u_{\rm ex}k}{T}\right)-1\right]^{-1}.
\end{equation}
Since the collisions between excitations conserve their total
momentum, the equilibrium is characterized by the velocity $u_{\rm
  ex}$, which is not necessarily equal to the velocity $u_0$
associated with the Fermi surface.

As discussed above, in the absence of backscattering there are four
conserved macroscopic characteristics of the fluid: the number of
particles, energy, momentum and $J$.  The hydrodynamic description of
the fluid is obtained by writing these conservation laws in the form
of continuity equations on the respective densities:
\begin{subequations}
  \label{eq:continuity}
  \begin{eqnarray}
    \label{eq:continuity_n}
    \partial_t n + \partial_x j&=&0,
\\
    \label{eq:continuity_energy}
    \partial_t \varepsilon + \partial_x j_\varepsilon&=&0,
\\
    \label{eq:continuity_p}
    \partial_t p + \partial_x j_p&=&0,
\\
    \label{eq:continuity_u_0}
        \partial_t u_0 + \partial_x j_{u_0}&=&0.
  \end{eqnarray}
\end{subequations}
Here $n$, $\varepsilon$, and $p$ are densities of particles, energy,
and momentum of the system, respectively.  Instead of density $J/L$ we
use the velocity $u_0$ defined by Eq.~(\ref{eq:u_0}).  The
corresponding currents $j$, $j_\varepsilon$, $j_p$, and $j_{u_0}$ are
yet to be determined. 

Below we only consider the regime of small deviation of the system
from thermal equilibrium, which will be described by two velocities
$u_0$ and $u_{\rm ex}$ and the deviations of densities $n$ and $s$ of
particles and entropy from mean values, $n-n_0$ and $s-s_0$.  We start
by evaluating $\varepsilon$ and $p$ in the leading order in these
small parameters.  At finite temperature, the dominant contribution to
the energy density $\varepsilon$ is due to the quasiparticle
excitations.  Substituting the occupation numbers (\ref{eq:Bose}) into
the last term in Eq.~(\ref{eq:E}), we obtain
\begin{equation}
  \label{eq:varepsilon}
    \varepsilon=\frac{\pi T^2}{6\hbar {\tilde v}} 
    =\frac{3\hbar}{2\pi}{\tilde v} s^2,
\qquad {\tilde v} = \left(\frac{1}{v_\rho}+\frac{1}{v_\sigma}\right)^{-1}.
\end{equation}
Here we applied the relation $\partial \varepsilon/\partial s=T$ to
find the entropy density $s=\pi T/3\hbar {\tilde v}$ and expressed
$\varepsilon$ in terms of $s$.  Combining
Eqs.~(\ref{eq:u_0})--(\ref{eq:Bose}) with (\ref{eq:P}), we find the
momentum density
\begin{equation}
  \label{eq:p}
  p=mnu_0+\frac{2\varepsilon}{v_2^2}(u_{\rm ex}-u_0),
\qquad  v_2 = \left(
                \frac{v_\rho^{-1}+v_\sigma^{-1}}{v_\rho^{-3}+v_\sigma^{-3}}
              \right)^{1/2}.
\end{equation}

Then, using Galilean invariance we immediately obtain the particle
current $j=p/m$ in the form
\begin{subequations}
\label{eq:currents}
\begin{equation}
  \label{eq:j}
  j=nu_0+\frac{2\varepsilon}{mv_2^2}(u_{\rm ex}-u_0).
\end{equation}
The remaining three currents can be obtained using the kinetic
equation for elementary excitations and accounting for the fact that
collisions do not change the number of particles, momentum, energy and
$J$.  The method was developed in the theory of superfluidity
\cite{khalatnikov_introduction_2000}.  When applied to the Luttinger
liquid, the results take the form
\begin{eqnarray*}
  j_\varepsilon &=& \sum_{\lambda=\rho,\sigma}
          \int\frac{dk}{h}
          N^\lambda_k
            \left[j\partial_n\epsilon_\lambda(k)
               +\epsilon_\lambda(k)
          \frac{\partial\epsilon_\lambda(k)}{\partial k}\right],
\\
  j_p &=& j_p^{(0)} 
         + \sum_{\lambda=\rho,\sigma} \int\frac{dk}{h} N^\lambda_k
          \left[n\partial_n\epsilon_\lambda(k)
               +k \frac{\partial\epsilon_\lambda(k)}{\partial
                 k}\right],
\\
  j_{u_0}&=& \frac{1}{m}
            \left[\mu^{(0)}
           + \sum_{\lambda=\rho,\sigma} \int\frac{dk}{h} N^\lambda_k
             \partial_n\epsilon_\lambda(k)
            \right].
\end{eqnarray*}
Here $j_p^{(0)}$ and $\mu^{(0)}$ are the pressure and chemical
potential, respectively, of the Luttinger liquid at $T=0$.  Using
Eqs.~(\ref{eq:epsilon}) and (\ref{eq:Bose}), to leading order in $u_0$
and $u_{\rm ex}$ we find
\begin{eqnarray}
  \label{eq:j_varepsilon}
  j_\varepsilon &=& \varepsilon\frac{\partial_n {\tilde v}}{{\tilde v}}j
                  +2\varepsilon u_{\rm ex},
\\
  \label{eq:j_p}
  j_p &=& j_p^{(0)}
          + \varepsilon\frac{\partial_n {(n\tilde v)}}{{\tilde v}},
\\
  \label{eq:j_u_0}
  j_{u_0} &=& \frac{\mu^{(0)}}{m} 
             + \varepsilon\frac{\partial_n {\tilde v}}{m{\tilde v}}.
\end{eqnarray}
\end{subequations}

We are now in a position to transform Eq.~(\ref{eq:continuity}) into a
set of four differential equations on four hydrodynamic parameters of
the fluid: $n$, $s$, $u_0$ and $u_{\rm ex}$.  Substituting
Eq.~(\ref{eq:j}) into (\ref{eq:continuity_n}), we find
\begin{subequations}
\begin{equation}
  \label{eq:dtnu}
  \partial_t n + n\,\partial_x u_0 
        + \frac{2\varepsilon}{mv_2^2}
          (\partial_x u_{\rm ex} - \partial_x u_0)
     =0.
\end{equation}
When substituting Eq.~(\ref{eq:varepsilon}) into
(\ref{eq:continuity_energy}), one should use the expression in terms
of the entropy density $s$ and keep in mind that ${\tilde v}$ is a
function of density $n$ that in turn depends on time.  Expressing the
resulting $\partial_t n$ with the aid of Eq.~(\ref{eq:continuity_n})
and using the expression (\ref{eq:j_varepsilon}) for $j_\varepsilon$,
we obtain
\begin{equation}
  \label{eq:continuity_s}
  \partial_t s + s\,\partial_x u_{\rm ex} = 0.
\end{equation}
This result has the form of the continuity equation expressing the
conservation of entropy, which holds to linear order in a deviation
from equilibrium.  Since the entropy is transported only by the gas of
excitations, one expects the entropy current in the form $j_s=su_{\rm
  ex}$, in agreement with Eq.~(\ref{eq:continuity_s}).

When substituting Eqs.~(\ref{eq:j_p}) and (\ref{eq:j_u_0}) into
(\ref{eq:continuity_p}) and (\ref{eq:continuity_u_0}) one must
evaluate the derivatives of the ground state chemical potential
$\mu^{(0)}$ and pressure $j_p^{(0)}$ with respect to the density.  The
chemical potential is easily obtained from the first term in
Eq.~(\ref{eq:E}), resulting in $\partial_n\mu^{(0)}=mv_\rho^2/n$.  The
derivative of the pressure is found using the thermodynamic relation
$\partial_n j_p^{(0)}=n \partial_n \mu^{(0)}=mv_\rho^2$.  Then
Eq.~(\ref{eq:continuity_p}) takes the form
\begin{eqnarray}
  \label{eq:dtp}
  &&\hspace{-2em}\partial_t u_0 
  +\frac{2\varepsilon}{mnv_2^2}(\partial_t u_{\rm ex}-\partial_t u_0)
\nonumber\\
  &&\hspace{-1em}+v_\rho^2\left[
            1+\varepsilon\frac{\partial_n^2(n\tilde v)}
                              {mv_\rho^2\tilde v}
           \right]
            \frac{\partial_x n}{n}
         +\frac{2\varepsilon}{mn}\frac{\partial_n(n\tilde v)}{\tilde v}
         \frac{\partial_x s}{s}=0.
\end{eqnarray}
To leading order at $T\to0$, substitution of Eq.~(\ref{eq:j_u_0}) into
(\ref{eq:continuity_u_0}) gives the same result, because in this limit
$p=mnu_0$.  Taking the difference of these two equations, which
accounts for the time dependence of the momentum of the gas of
excitations, we arrive at
\begin{equation}
  \label{eq:dtp_ex}
  \partial_t u_{\rm ex}-\partial_t u_{0}
    +v_2^2\frac{n\partial_n\tilde v}{\tilde v}\frac{\partial_x n}{n}
    +v_2^2\frac{\partial_x s}{s}=0.
\end{equation}
\label{eq:hydrodynamic_equations}
\end{subequations}

To study the propagation of collective modes in one-dimensional
liquids, we now solve the system of equations
(\ref{eq:hydrodynamic_equations}).  In the low-temperature limit one
can set $\varepsilon=0$ in Eqs.~(\ref{eq:dtnu}) and (\ref{eq:dtp}).
One easily finds two propagating-wave solutions proportional to
$e^{-i\omega t+iqx}$.  First, Eqs.~(\ref{eq:dtnu}) and (\ref{eq:dtp})
give rise to a phonon-like mode with the spectrum $\omega=v_\rho |q|$.
This mode is determined by the dynamics of the variables $n$ and
$u_0$, describing the waves of particle density.  Due to the presence
of mixing terms in Eq.~(\ref{eq:dtp_ex}), the phonon is accompanied by
the oscillation of entropy density $s$ and velocity of the gas of
excitations $u_{\rm ex}$.

Second, there is a solution with the spectrum $\omega=v_2|q|$ that
describes waves of $s$ and $u_{\rm ex}$, whereas $n=n_0$ and $u_0=\rm
const$.  This wave of entropy is fully analogous to the second sound
in superfluid $^4$He.  The existence of the second sound in a system
of one-dimensional fermions with repulsive interactions is the main
result of this paper.

Our discussion so far assumed that the frequencies of interest are in
the range (\ref{eq:frequency_range}).  In other words, we set
$\tau_{\rm ex}=0$ and $\tau=\infty$.  We shall now relax the latter
condition, i.e., assume a large but finite $\tau$ and extend our
treatment to frequencies $\omega\lesssim\tau^{-1}$.  In this regime
one must account for the backscattering processes studied in
Refs.~\cite{micklitz_transport_2010, matveev_equilibration_2010,
  matveev_equilibration_2012, matveev_scattering_2014}.  Due to the
slow rate of these processes, they do not affect the equilibrium form
of the distribution function (\ref{eq:Bose}).  As a result the state
of the system is still described by parameters $n$, $T$, $u_0$, and
$u_{\rm ex}$, but because of the backscattering processes the two
velocities relax toward each other as
\begin{equation}
  \label{eq:tau_definition}
  \frac{d}{dt}(u_{\rm ex}-u_0^{}) = -\frac{u_{\rm ex}-u_0^{}}{\tau}.
\end{equation}
It is important to point out that this relaxation does not affect the
expressions (\ref{eq:currents}) for the currents and does not violate
the conservation laws for the number of particles, energy, and
momentum of the system.

In our hydrodynamic description of the one-dimensional system, the
first three of the four equations (\ref{eq:continuity}), and,
respectively, (\ref{eq:hydrodynamic_equations}), express these three
conservation laws and thus remain unchanged.  The right-hand side of
Eq.~(\ref{eq:continuity_u_0}), becomes $du_0/dt$, which is found by
applying conservation of momentum condition $dp/dt=0$ to
Eq.~(\ref{eq:p}) and using (\ref{eq:tau_definition}).  After that we
recover Eq.~(\ref{eq:dtp_ex}) with a simple modification $\partial_t
\to \partial_t+\tau^{-1}$.

This modification of the hydrodynamic equations
(\ref{eq:hydrodynamic_equations}) strongly affects the second sound
mode at $\omega\lesssim\tau^{-1}$.  To first approximation we take
$\varepsilon/nmv_2^2\to0$ and obtain the frequency of the second sound
in the form
\begin{equation}
  \label{eq:second_sound_modified}
  \omega=\sqrt{(v_2 q)^2-(2\tau)^{-2}}-i(2\tau)^{-1}.
\end{equation}
At $v_2|q|>(2\tau)^{-1}$ the frequency is reduced, and more
importantly, the second sound decays with the rate $(2\tau)^{-1}$.  No
wavelike solution exists at $v_2|q|<(2\tau)^{-1}$.  Heat propagation
over long distances is diffusive: $\omega=-i(v_2^2\tau)q^2$ at
$q\to0$.  As a result, the system has a large, but finite thermal
conductivity $\kappa$ obtained by multiplying the diffusion
coefficient $v_2^2\tau$ by the specific heat $\partial
\varepsilon/\partial T$,
\begin{equation}
  \label{eq:kappa}
  \kappa=\frac{\pi T v_2^2 \tau}{3\hbar\tilde v}.
\end{equation}
Alternatively, the thermal conductivity can be obtained directly from
the modified Eq.~(\ref{eq:dtp_ex}).  Replacing $\partial_t
\to \partial_t+\tau^{-1}$ and considering long time scales gives
Eq.~(\ref{eq:dtp_ex}) with $\partial_t \to \tau^{-1}$.  At
$\varepsilon/nmv_2^2\to0$ the gas of excitations does not affect
particle density $n$ and velocity $u_0$, see Eqs.~(\ref{eq:dtnu}) and
(\ref{eq:dtp}).  Assuming $n=\rm const$ and $u_0=\rm 0$ in
Eq.~(\ref{eq:dtp_ex}), one finds $u_{\rm
  ex}=-v_2^2\tau(\partial_xs)/s$.  Substituting this result into the
expression $j_Q=Tsu_{\rm ex}$ for the heat current, we obtain
$j_Q=-Tv_2^2\tau\partial_xs$.  Using our earlier result for the
entropy density $s=\pi T/3\hbar {\tilde v}$, we obtain
$j_Q=-\kappa\partial_xT$ with $\kappa$ given by Eq.~(\ref{eq:kappa}).

To find the effect of a finite backscattering rate on the first sound,
one should solve the set of equations
(\ref{eq:hydrodynamic_equations}) in first order in the small parameter
$\varepsilon/nmv_2^2$.  At small $q$ we find
\begin{equation}
  \label{eq:omega_first_sound}
  \omega=\left[v_\rho+\frac{\pi T^2\partial_n^2(n^2\tilde v)}{12\hbar mn
    v_\rho\tilde v^2}\right] q
  -i\frac{\kappa T}{2mnv_\rho^2}\left[\frac{\partial_n(n\tilde
      v)}{\tilde v}\right]^2q^2.
\end{equation}
This result demonstrates that at $\omega\tau\to0$ the first sound mode
becomes the ordinary thermodynamic sound.  In particular, the first
term in Eq.~(\ref{eq:omega_first_sound}) contains a correction to the
sound velocity, which simply accounts for the temperature dependence
of the adiabatic compressibility of the one-dimensional quantum
liquid.  The second term is imaginary and thus describes attenuation
of the sound mode.  Indeed, in any medium thermal conductivity gives
rise to the absorption of sound.  We have verified that the resulting
absorption rate \cite{landau_fluid_2013} is consistent with the second
term in Eq.~(\ref{eq:omega_first_sound}).

To summarize, we have studied collective excitations of a system of
one-dimensional spin-$\frac12$ fermions at a low temperature based on
a two-fluid hydrodynamic description of the system.  In contrast to
liquid $^4$He, there is no superfluid condensate in our case.  The
two-fluid nature of the system can be understood as follows.  We apply
the Luttinger liquid theory to small sections of the one-dimensional
system.  The state of each section is described by two sets of
variables: the occupation numbers of the elementary excitations, and
the zero modes $N$ and $J$.  In addition, we keep in mind that the
excitations equilibrate with each other at the rather short time scale
$\tau_{\rm ex}$, whereas their equilibration with the zero modes
happens at the much longer scale $\tau$.  Thus in the frequency range
(\ref{eq:frequency_range}) the system consists of two components.  The
excitations form a gas, analogous to the normal component of
superfluid $^4$He, whereas the position- and time-dependent values of
densities of $N$ and $J$ describe a second liquid, similar to the
superfluid component of $^4$He.

Our main result is that in addition to the well-understood acoustic
charge and spin excitation modes propagating at velocities $v_\rho$
and $v_\sigma$, there is a second sound mode propagating at velocity
$v_2$ given by Eq.~(\ref{eq:p}).  This mode describes the waves of
entropy; its decay is small for frequencies in the range
(\ref{eq:frequency_range}).  In contrast to superfluid $^4$He, at
$\omega\ll \tau^{-1}$ the second sound disappears, and the heat
transport becomes diffusive.  Another system where the second sound
exists in a finite frequency range is dielectric crystal
\cite{enz_two-fluid_1974, gurevich_transport_1986}.

Our treatment can be applied to other one-dimensional systems at low
temperatures, such as a system of bosons or spin-polarized fermions.
The absence of spin excitations in these systems can be accounted for
by taking the limit $v_\sigma\to\infty$ in our formulas.  In this case
the velocities of the first and second sound modes are both equal to
$v_\rho$ in the limit $T\to 0$.  It is worth mentioning that at
$v_\sigma\to\infty$ our expression (\ref{eq:kappa}) for thermal
conductivity recovers the result for the spinless one-dimensional
system obtained in Ref.~\cite{degottardi_electrical_2015}.  Another
important example is that of spin-$\frac12$ fermions with attractive
interactions.  In this case the energy spectrum of the spin
excitations has a finite gap $\Delta$ at $T=0$
\cite{giamarchi_quantum_2004}.  Our two-mode description still applies
at $\Delta\ll T$.  In the opposite limit, $\Delta\gg T$, the spin
excitations are frozen out, and the adaptation of the theory to the
spinless case, as described above, should be made.

The existence of the second sound mode means that the heat propagation
in the one-dimensional system is ballistic at sufficiently high
frequencies $\omega\gg\tau^{-1}$, whereas the usual diffusive heat
transport is restored at $\omega\ll\tau^{-1}$.  Experimentally, such a
frequency dependence of thermal transport may be observed in long
ballistic quantum wires, such as those obtained by the cleaved-edge
overgrowth technique \cite{yacoby_nonuniversal_1996,
  scheller_possible_2014}.  A time-dependent temperature difference
across the wire can be achieved by driving ac current through one of
the leads, cf. Ref.~\cite{staring_coulomb-blockade_1993}.

A direct observation of both first and second sound was recently
reported in a system of $^6$Li atoms in elongated trap
\cite{sidorenkov_second_2013}.  In this experiment the system was
three-dimensional, and superfluidity was achieved by tuning
interactions to resonance by a magnetic field.  In order to observe
the second sound discussed in this paper, one can replace the trap in
Ref.~\cite{sidorenkov_second_2013} with an array of narrow traps that
are in the one-dimensional regime \cite{kinoshita_observation_2004,
  moritz_confinement_2005}.

\begin{acknowledgments}

  The authors are grateful to Subhadeep Gupta for helpful
  discussions. Work at Argonne National Laboratory was supported by
  the U.S. Department of Energy, Office of Science, Materials Sciences
  and Engineering Division.  Work at the University of Washington was
  supported by the U.S.  Department of Energy Office of Science, Basic
  Energy Sciences under Award No. DE-FG02-07ER46452.

\end{acknowledgments}


\begin{thebibliography}{34}
\bibitem [{foo()}]{footnote}%
  \BibitemOpen
  \href@noop {} {}\bibinfo {note} {A more detailed microscopic theory is
  required for the evaluation of the relaxation times $\tau$ and $\tau_{ex}$,
  which is not part of this work.}\BibitemShut {Stop}%
\end{thebibliography}
\end{document}